\documentclass[twocolumn,pre]{revtex4}
\usepackage{graphicx}
\usepackage{times} 
\usepackage{palatino} 
\usepackage{newcent} 
\usepackage{bookman}

\newcommand{\half}{\frac{1}{2}}

\newcommand{\bi}{\begin{itemize}}
\newcommand{\ei}{\end{itemize}}
\newcommand{\be}{\begin{equation}}
\newcommand{\ee}{\end{equation}}

\newcommand{\npclass}{{NP}}

\newcommand{\kcol}{{k-col}}

\begin{document}

\title{From N\'eel to NPC: Colouring Small Worlds}
\date{\today}
\author{Pontus Svenson \\
tfkps@fy.chalmers.se}
\affiliation{Institute for Theoretical Physics \\
Chalmers University of Technology and G\"oteborg University \\
SE-412 96 Gothenburg, Sweden}

\begin{abstract}
In this note, we present results for the colouring problem
on small world graphs created by rewiring square, triangular, and two kinds
of cubic (with coordination numbers 5 and 6) lattices.
As the rewiring parameter $p$ tends to 1, we find the expected crossover
to the behaviour of random graphs with corresponding connectivity.
However, for the cubic lattices 
there is a region near $p=0$ for which the graphs are colourable.
This could in principle be used as an additional heuristic for solving real
world colouring or scheduling problems. Small worlds with connectivity 5
and $p\sim 0.1$ provide an interesting ensemble of graphs whose
colourability is hard to determine.
For square lattices, we get good data collapse plotting the fraction
of colourable graphs against the rescaled parameter 
parameter $p N^{-\nu}$ with $\nu = 1.35$. No such collapse
can be obtained for the data from lattices with coordination number 5 or
6.
\end{abstract}

\maketitle

Several \npclass-complete~\cite{papadimitriou}
constraint satisfaction problems have recently been
found to show threshold phenomena. As some parameter is varied,
the fraction of problems than can be solved suddenly changes
from 1 to 0 (e.g.,~\cite{cheesemankanefskytaylor,hogghubermanwilliams}).
A related transition occurs for the difficulty of determining
whether a problem instance
is solvable. Far from the transition, this is almost always
easy, while computationally hard problems are clustered near
the critical value of the parameter.
These transitions have been compared to phase transition appearing
in physical systems, and has attracted much interest
from physicists who have used techniques from statistical physics such as
finite-size scaling~\cite{kirkpatrickselman} and replica symmetry
breaking~\cite{monassonzecchina} to study it. Most of the
work by physicists has so far been concentrated on the 
boolean satisfiability problem, though it should be noted that there
has been much analytical work on the number partitioning
problem~\cite{mertens}.

One important \npclass-complete problem is the k-colouring problem 
(\kcol). It consists of colouring all the nodes
of a graph using k colours
in such a way that no edge joins two nodes of the same
colour. This problem is particularly interesting for physicists,
since it corresponds to finding the zero temperature
ground state of an antiferromagnetic
k-state Potts model. Graph colouring also has many real-world
applications, including scheduling and register-allocation problems.
For the \kcol\ problem on random graphs with average
degree $\gamma$, Achlioptas and Friedgut~\cite{achlioptasfriedgut} 
have shown rigorously that there is a $\gamma_c$ where the
fraction of colourable graphs jumps from 1 to 0. Even though there
is currently no exact analytical result for the value of $\gamma_c$,
Achlioptas~\cite{achlioptas:thesis} has bound it rigorously
by $3.84 < \gamma_c < 5.05$.

Small world graphs have recently been introduced in an attempt
to capture the topological properties of real graphs. Regular
lattices have long average distance between nodes, but show
a high degree of clustering (i.e., if we remove a node from
the graph, its neighbours will still have short paths between them).
On the other hand, random graphs~\cite{bollobas} have
no clustering and small average distances. Watts and 
Strogatz~\cite{wattsstrogatz:nature,wattsbook} have therefore
proposed a new graph model. In their model, we start from a
regular graph and apply a rewiring process to each edge
in it: with probability $p$, the edge is replaced
by a short-cut, linking one of the original nodes 
to a randomly selected other
node in the graph. The total number of edges (and hence
the mean connectivity) is thus conserved in the small world model.
In Watts and Strogatz's original model,
a ring lattice where each node is connected to 
its $2k$ nearest neighbours is used as a starting point;
most of the work done on small worlds use
this model too. For recent reviews on this and
related graph models, 
see~\cite{newmanreview,newmanstrogatzwatts2000}.

Here we use other lattices as starting points. The motivation for
this is to determine the influence of the small world properties
of the graphs on the phase transition. Since $\gamma=4$ for the
2D square lattice and $\gamma=6$ for the 2D triangular
and 3D simple cubic lattices, we know that for $p=1$ they should
display completely different behaviour 
(at least in the thermodynamic limit of $N\to \infty$).
For $p=0$, all lattices used in this paper are colourable using 3 colours.
(In physics the repeating pattern of alternating colours used for this
is called a N\'{e}el state.)

The hardness of
colouring small-world graphs has previously been investigated by
Walsh~\cite{walsh:searchsmall}, who used the original linear model of 
Waltz and Strogatz. 
In addition to studying some other constraint satisfaction
problems on small world graphs, Walsh showed that the cost to prove that
a small world graph is colourable has an ``easy-hard-easy'' behaviour
as a function of $p$.
The results presented here are different, since
the rewiring procedure starts from other lattices.
In particular, since
it is known~\cite{hogg:asmcsp} that 
\kcol\ on random graphs displays a ``phase-transition'' as
the connectivity $\gamma$ passes a threshold value $\gamma_c \approx 4.6$,
it is worthwhile to investigate whether there are differences when
starting from lattices with connectivity 4, 5, and 6.

Another motivation for using other lattices as starting points
is an interpretation of \kcol\ as a toy model of a complex system
that displays frustration. Such models need to be placed on more
realistic graphs.
For example, networks that model the interactions between agents
occupying different parts of a land area and competing for
some common resource are best modeled
using the square lattice as a starting point.

To colour the graphs, a backtrack search program based on
the Brelaz heuristic~\cite{brelaz} was used. Simulations were performed
for the 2D square (connectivity $\gamma=4$) and triangular ($\gamma=6$)
lattices as well as two different 3D simple cubic lattices, with
$\gamma=5$ and 6. The cubic lattice with $\gamma=5$ was obtained from
the simple cubic lattice by
deleting every second vertical edge in a chess-board pattern.

The linear size of the square lattice was varied from 4
to 7, while the experiments on the cubic lattices used $L=3$, 4, and 5.
Triangular lattices of linear size 6 and 9 were also tested. For
each $p$ and lattice, an average was performed over up to $10^7$
different graphs. 
A considerable smaller number of graphs were used for the large $L$ 
and all the $\gamma=5$ runs. This was necessary because of the 
``easy-hard-easy''-transition and because of the exponential 
time-complexity of the Brelaz algorithm. The number of nodes
visited in the search tree was used as a measure of the cost of
colouring a graph or determining that no solution exists. 
The random number generator used to generate most of the 
graphs is the Mitchell-Moore generator (see, e.g.,~\cite{knuth2}). 
Similar results were obtained in test runs
using the standard C library's {\tt drand48()} generator.

\begin{figure}
\centering
\leavevmode
\includegraphics[width=.75 \columnwidth]{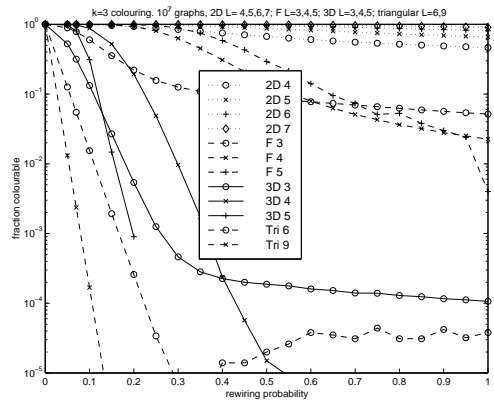}
\vspace*{2mm}
\caption{This semilogarithmic plot shows the fraction of colourable graphs as a
function of $p$ for various starting lattices. In the legend,
``2D'', ``F'', ``3D'', and ``Tri'' denote the square, 
$\gamma=5$ cubic, normal cubic, and triangular starting lattices.
The numbers in the legend are
the linear size of the lattice. Data were obtained by averaging over 
from 1000 (for some of the ``F'' data) to $10^7$ different graphs.
The differences between the data for graphs with 
$\gamma=4$, 5, and 6 is clearly seen. See also figures~\ref{squarefivesemi}
and~\ref{fivecubesemi} for detailed comparisons of the different $\gamma$'s.}
\label{alldataloglin}
\end{figure}

Figure~\ref{alldataloglin} shows the results for the fraction
of colourable graphs for all different start lattices.  
Notice that this and the other
plots are in a semilogarithmic scale.
Plot legends describe the lattice and
its linear size. The symbol ``F'' denotes the $\gamma=5$ cubic lattice; thus
``F 4'' denotes a system of size $N=4^3=64$ variables.

\begin{figure}
\centering
\leavevmode
\includegraphics[width=.75 \columnwidth]{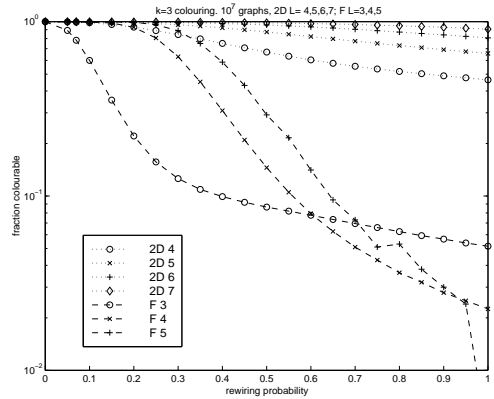}
\vspace*{2mm}
\caption{This shows the difference between the square lattice (top curves)
and
$\gamma=5$ cubic (bottom curves) data. 
(Same data as in figure~\ref{alldataloglin})}
\label{squarefivesemi}
\end{figure}

\begin{figure}
\centering
\leavevmode
\includegraphics[width=.75 \columnwidth]{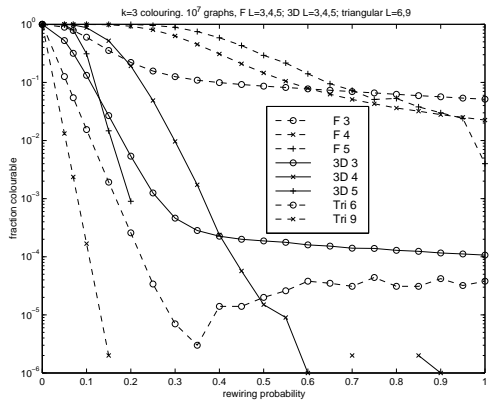}
\vspace*{2mm}
\caption{This shows the difference between the
$\gamma=5$ cubic (top curves) 
and $\gamma=6$ cubic and triangular lattice (bottom curves)
data. (Same data as in figure~\ref{alldataloglin})}
\label{fivecubesemi}
\end{figure}

In figure~\ref{squarefivesemi} we compare the data for the square and
$\gamma=5$ cubic lattices, while figure~\ref{fivecubesemi} shows
the $\gamma=5$ and $\gamma=6$ data from the runs on the
triangular and cubic lattices. The phase transition in solvability
can be seen quite clearly in these figures. But note that even for rather
large $p$ and $N$, there is still a rather high probability that a $\gamma=5$
graph can be coloured. In fact, the
appearance of the $L=3$, 4, and 5 curves seems to indicate that there
could be a cross-over at a finite $p$. Unfortunately, the exponential
time requirements of the Brelaz algorithm currently prevents us 
from investigating this in more detail numerically.

\begin{figure}
\centering
\leavevmode
\includegraphics[width=.75 \columnwidth]{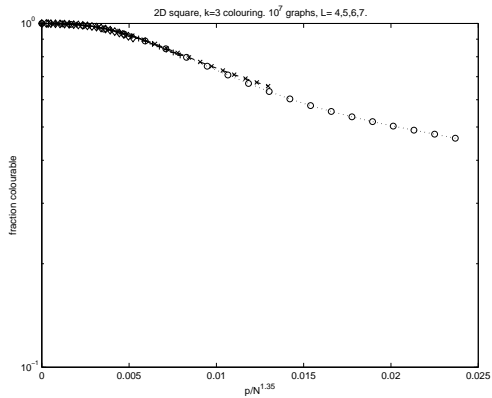}
\vspace*{2mm}
\caption{For the data from square ($\gamma=4$) lattices, a
good collapse can be obtained by plotting the fraction
of colourable graphs as a function of $p^{-\nu}$ with $\nu=3.5$.}
\label{squarecollapse}
\end{figure}

As seen in figure~\ref{squarecollapse}, good data collapse
can be obtained for the square lattices by plotting the solvable fraction as a function
of $p N^{-\nu}$ with $\nu = 1.35$. No such collapse can however
be obtained for the cubic or triangular data.

\begin{figure}
\centering
\leavevmode
\includegraphics[width=.75 \columnwidth]{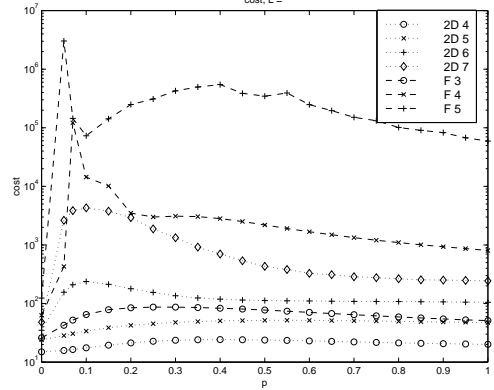}
\vspace*{2mm}
\caption{The average cost of determining whether a graph can be coloured
or not is also interesting. Here we compare the cost for $\gamma=4$ and
$\gamma=5$ as functions of $p$. Note that the cost depends
more on the number of variables in the problem than on the
coordination number of the graphs. Large system sizes are necessary to
see the ``easy-hard-easy'' transition.}
\label{squarefivecost}
\end{figure}

\begin{figure}
\centering
\leavevmode
\includegraphics[width=.75 \columnwidth]{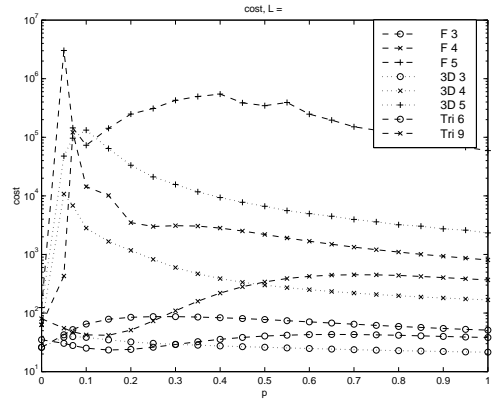}
\vspace*{2mm}
\caption{As in figure~\ref{squarefivecost}, here we compare the costs for
$\gamma=5$ and $\gamma=6$ as functions of $p$. The odd behaviour of
the data for triangular lattices is due to finite size effects.}
\label{fivecubecost}
\end{figure}

In figures~\ref{squarefivecost} and~\ref{fivecubecost}, we show
the cost (measured as the number of nodes visited in the search-tree)
for the runs performed. The
$\gamma=5$ graphs prove to be much harder than the others. This is 
natural since they are closer to the phase transition for random
graphs. For the other data, the cost is essentially determined
by the number of variables in the problem, independent of 
which lattice was used as a starting point.
The ``easy-hard-easy''
pattern which can be seen for some of the data in these figures as a function of $p$ is
similar to that of \kcol\ on random graphs.

In conclusion,
we studied graph colouring on small world graphs produced
by starting with regular lattices of coordination number 4, 5, and 6. The
hardness patterns found are similar to those for the
original small world model. More interestingly, we found that for
finite system sizes, there is a range of $p$'s for which the graphs
are still colourable even for $\gamma=5$ and 6. We found a new class
of problems that are hard for the Brelaz algorithm and could be used
as a benchmark for new algorithms.
Investigating constraint satisfaction problems on small world graphs and
other random graph ensembles with topologies different from that of the
standard model ${\cal{G}}(N,\half \gamma N)$ is important because
these graphs might better capture the appearance of real world
instances of the problems. For the same reason, it would be interesting
to look at other constraint satisfaction problems (such as the
vertex-cover problem) on such graphs.

\end{document}